\documentclass[conference]{IEEEtran}
\usepackage{cite}

\ifCLASSINFOpdf
   \usepackage[pdftex]{graphicx}
\else
   \usepackage[dvips]{graphicx}
\fi

\usepackage[cmex10]{amsmath}
\usepackage{amsfonts,amssymb}

\usepackage{color}

\newtheorem{proposition}{Proposition}
\newtheorem{algorithm}{Algorithm}
\newcommand{\prox}[2]{\mathrm{prox}_{#1} \left(#2\right)}
\DeclareMathOperator*{\argmin}{arg\,min}
\DeclareMathOperator*{\argmax}{arg\,max}

\renewcommand{\IEEEQED}{\IEEEQEDopen}

\begin{document}
%
\title{Multiuser Detection by MAP Estimation \\ with Sum-of-Absolute-Values Relaxation}

\IEEEoverridecommandlockouts

\author{\IEEEauthorblockN{Hampei Sasahara, Kazunori Hayashi, and Masaaki Nagahara}
\IEEEauthorblockA{Graduate School of Informatics, Kyoto University, Japan \\
Email: sasahara.h@acs.i.kyoto-u.ac.jp, kazunori@i.kyoto-u.ac.jp, nagahara@ieee.org
\thanks{This research was supported in part by JSPS KAKENHI Grant Numbers
15H02668, 15K14006, 26120521, 15K06064, and 15H02252.}}
}


%


\maketitle

\begin{abstract}
In this article, we consider multiuser detection that copes with multiple access interference 
caused in star-topology machine-to-machine (M2M) communications.
We assume that the transmitted signals are discrete-valued (e.g. binary signals
taking values of $\pm 1$),
which is taken into account as prior information in detection.
We formulate the detection problem as the maximum \textit{a posteriori} (MAP) estimation,
which is relaxed to a convex optimization called the sum-of-absolute-values (SOAV) optimization.
The SOAV optimization can be efficiently solved by
a proximal splitting algorithm, for which we give the proximity operator in a closed form.
Numerical simulations are shown to illustrate the effectiveness of the proposed approach
compared with the linear minimum mean-square-error (LMMSE) and
the least absolute shrinkage and selection operator (LASSO) methods.
\end{abstract}


%
\IEEEpeerreviewmaketitle

\section{Introduction}
\label{sec:intro}

Machine-to-machine (M2M) communications, as part of the Internet of Things (IoT),
has drawn a great deal of interests from scientific and engineering communities.
M2M is considered as the next technology revolution after the Internet,
and has attracted more and more attention over the last few years;
see survey papers~\cite{AtzIerMor10,M2M} and references therein.

For typical M2M communications, the data rate can be quite low,
and the code division multiple access (CDMA) is a possible candidate
for the narrow-band M2M communications~\cite{Bockelmann2013}.
In such M2M communications, the number of nodes is often very large,
and hence multiuser detection is essential to cope with multiple access interference~\cite{Verdu}.
For this problem, various schemes have been proposed for CDMA systems such as
the linear minimum mean-square-error (LMMSE) method~\cite{Xie1990} and 
the maximum likelihood (ML) method~\cite{Etten1976,Verdu1986}.
The LMMSE method, however, does not demonstrate 
sufficient performance in many cases,
and the ML method in general requires a heavy computational burden
for large scale systems;
see~\cite{Moshavi1996} for details.

To overcome these drawbacks in the LMMSE and ML methods,
the characteristic of the transmitted signals in M2M should be
appropriately utilized as \emph{prior information} in design.
For this purpose, the notion of \emph{sparsity}~\cite{CompHayashi} has been adapted to M2M communications.
In many applications of M2M communications, transmitted signals can be modeled
as sparse signals in the time domain or the frequency domain.
For example, in security or guard services, an alert signal of
intrusion is sent to a control center~\cite{Ahn2010}, which can happen very rarely,
and hence the alert signal is sufficiently sparse in time.

Multiuser detection methods with the prior information of sparsity have been investigated
very recently by
a greedy algorithm called the orthogonal matching pursuit (OMP)~\cite{Shim2012}
and an $\ell^1$-based convex optimization method called
the least absolute shrinkage and selection operator (LASSO)~\cite{Zhu2011}.
In particular, in~\cite{Zhu2011},
the sparsity-exploiting sphere decoding method that takes candidates 
of transmitted symbols into account has been proposed.
Although these approaches can achieve better performance 
compared with the LMMSE and ML methods for sparse signals,
it cannot be applied to \emph{discrete-valued signals}
(e.g. binary signals taking values of $\pm 1$), 
which are also often used in M2M communications, but are not necessarily sparse.

\begin{figure}[t]
\centering
\includegraphics[width = .95\linewidth]{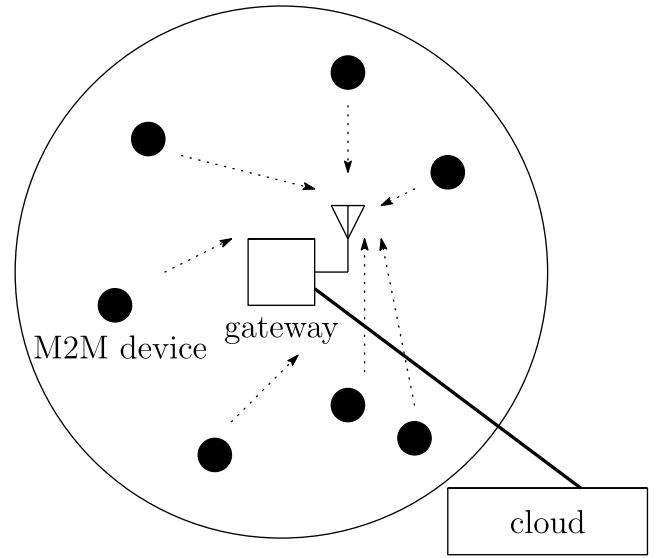}
\caption{M2M communication with a centralized structure.}
\label{m2m}
\end{figure}

In this article, we propose a new multiuser detection method taking account of
discreteness as prior information of transmitted signals
in the uplink M2M communications with a centralized structure
(or a \textit{star-topology}), as shown in Fig.~\ref{m2m}.
We formulate the multiuser detection problem as a maximum \textit{a posteriori} (MAP) estimation problem,
which is described as a combinatorial minimization problem with a sum of $\ell^0$ pseudo-norms.
In general, this optimization problem cannot be solved in polynomial time,
and we replace the sum of $\ell^0$ pseudo-norms with that of the $\ell^1$ norms
based on the idea of \textit{sum-of-absolute-values} (SOAV) optimization proposed in~\cite{DSR}.
The optimization problem is naturally described as linear programming,
which is still hard to solve for very large-scale problems,
and we adapt the method of \emph{proximal splitting}~\cite{Fixed-Point} to this problem.
This method requires the closed form of the proximity operator, which we give in this article.
Simulation results show that the proposed method, called MAP-SOAV,
shows superior performance to LMMSE~\cite{Xie1990} and LASSO~\cite{Zhu2011}.

The remainder of this article is as follows:
Section \ref{sec:sys} gives the system model considered in this article.
Section \ref{sec:det} reviews the LMMSE and LASSO estimations.
In Section \ref{sec:alg}, we propose the MAP-SOAV method 
and introduce an algorithm solving the derived optimization problem based on a proximal splitting method.
Moreover, the associated proximity operator is given as a closed form.
Numerical simulations are shown in Section \ref{sec:sim} to illustrate the effectiveness of the proposed method.
Section \ref{sec:conc} draws conclusions.

\section{System Model}
\label{sec:sys}
We consider synchronous multiuser communication.
Let the number of users be $N$.
A signature waveform $s_n(t)$ is assigned to the $n$th user $(1\leq n \leq N, n \in \mathbb{N})$.
$t$ denotes continuous time and it is assumed that $s_n(t)$ has the finite support $[0,T]$.
The transmission power is normalized to $1$.
Let $b_{n,k} \in \mathcal{A}$ be the $k$th transmission symbol of the $n$th user, where $\mathcal{A}$ stands for the finite set of the candidates of symbols.
$b_{n,k} = 0$ means that the $n$th user is not active on the $k$th transmission timing.
Note that we consider only real valued signal in this article for the sake of simplicity,
but the following statement can be extended to a complex valued case.

\begin{figure}[t]
\centering
\includegraphics[width = .95\linewidth]{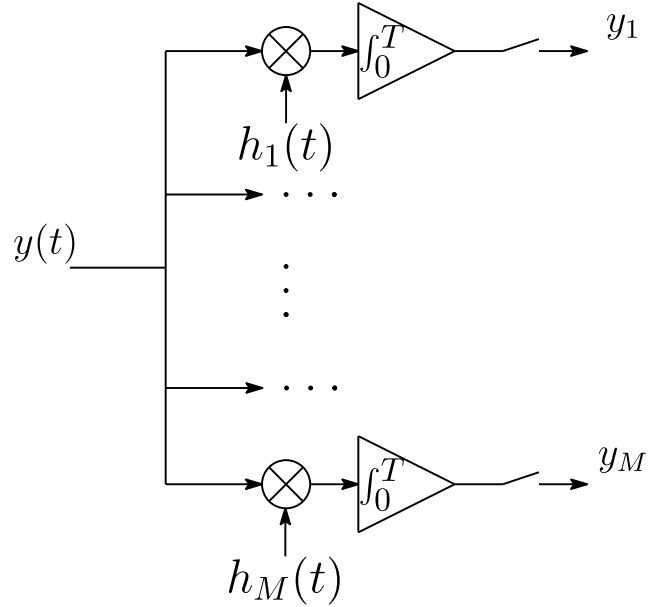}
\caption{Filter bank at the receiver.}
\label{FB}
\end{figure}

The received signal $y(t)$ through an additive white Gaussian noise (AWGN) channel is given by
\[
 y(t) = \sum_{k}\sum_{n=1}^N a_n b_{n,k} s_n(t-kT) + w(t),
\]
where $a_n \in \mathbb{R}$ is the $n$th user's channel gain and $w(t)$ is white Gaussian noise with zero mean and variance of $\sigma_w^2$.
At the receiver, we use the filter bank consisting of $M$ filters shown in Fig.~\ref{FB},
where $h_m(t)\ (1\leq m \leq M)$ is the impulse response of the $m$th filter.
We now consider the symbols for the case that $k=0$.
Let $y_m$ be the output of the $m$th filter
and $y_m$ is given by
\[
 y_m = \sum_{n=1}^N a_n b_n \int_0^T s_n(t)h_m(T-t)dt + \int_0^T w(t)h_m(T-t) dt.
\]
We define
\[
 \begin{array}{c}
 \displaystyle{\tilde{s}_{mn} \triangleq \int_0^T s_n(t) h_m(T-t)dt,} \\
 \displaystyle{\tilde{w}_m \triangleq \int_0^T w(t) h_m(T-t)dt.} \\
 \end{array}
\]
Then, we get
\begin{equation}
 \tilde{y} = \tilde{S} A b + \tilde{w},
 \label{equ1}
\end{equation}
where
\[
 \begin{array}{c}
 \tilde{y} \triangleq [y_1, \ldots, y_M]^{\top}, \\
 \tilde{S} \triangleq \left[
 \begin{array}{ccc}
 s_{11} & \ldots & s_{1N} \\
 \vdots & & \vdots \\
 s_{M1} & \ldots & s_{MN} \\
 \end{array}\right], \\
 A \triangleq \mathrm{diag}(a_n), \\
 b \triangleq [b_1, \ldots, b_N]^{\top}, \\
 \tilde{w} \triangleq [\tilde{w_1},\ldots,\tilde{w_M}]^{\top}, \\
 \end{array}
\]
and $\mathrm{diag}(a_n)$ represents the diagonal matrix whose diagonal elements are $a_1,\ldots,a_N$.
Let $\Sigma_{\tilde{w}}$ be the variance-covariance matrix of $\tilde{w}$.
Then, the ($i,j$)th component of $\Sigma_{\tilde{w}}$ is given by
\[
 \begin{array}{ccl} \vspace{.2em}
 (\Sigma_{\tilde{w}})_{ij} & = & \displaystyle{\mathbb{E}\left[ \int_0^T w(t) h_i(T-t) dt \int_0^T w(t)h_j(T-t) dt \right]} \\
 & = & \displaystyle{\int_0^T \int_0^T h_i(T-t) h_j(T-u) \mathbb{E} \left[ w(t)w(u)\right] dudt} \\ \vspace{.2em}
 & = & \displaystyle{\int_0^T \int_0^T h_i(T-t) h_j(T-u) \sigma_w^2 \delta(t-u) dudt} \\
 & = & \displaystyle{\sigma_w^2 \int_0^T h_i(T-t) h_j(T-t) dt} \\
 & = & \displaystyle{\sigma_w^2 \int_0^T h_i(t)h_j(t) dt,} \\
 \end{array}
\]
where $\delta(\cdot)$ denotes the delta distribution.
Hence, we obtain
\[
 \Sigma_{\tilde{w}} = \sigma_w^2 H,
\]
where the ($i,j$)th component of $H$ is defined by $\int_0^T h_i(t) h_j(t) dt$.
Finally, multiplying $H^{-1}$ to the left hand side of (\ref{equ1}), we get
\begin{equation}
 y = SAb+w,
\label{eq2}
\end{equation}
where
\[
 \begin{array}{c}
 y \triangleq H^{-1}\tilde{y}, \\
 S \triangleq H^{-1}\tilde{S}, \\
 w \triangleq H^{-1}\tilde{w}. \\
 \end{array}
\]
Note that the variance-covariance matrix of $w$ is $\sigma_w^2 I$, where $I$ is the unit matrix.
The multiuser detection problem considered in this article is to estimate $b$ from $y$ with the relationship (\ref{eq2}).

\section{Conventional Schemes}
\label{sec:det}

In this section, we briefly review the LMMSE method~\cite{Verdu} and the LASSO method~\cite{LASSO}.

\subsection{LMMSE estimation}
The transmitted signal $b$ is estimated with the LMMSE weight matrix $W$ by
\[
 b_{\mathrm{LMMSE}} = Wy.
\]
Let $\Sigma_b$ be the covariance matrix of $b$.
$W$ is calculated as follows~\cite{Proakis}:
\[
 W = \Sigma_b AS^{\top} (SA \Sigma_b A S^{\top} + \sigma_w^2 I)^{-1}.
\]
We assume that $\rho$ $(0\leq \rho < 1)$ represents the non-active rate and obtain
\[
 \begin{array}{ccl}
 \Sigma_b & = & \mathbb{E}[bb^{\top}] \\
 & = & (1-\rho)I. \\
 \end{array}
\]
Thus,
\[
 W = (1-\rho) AS^{\top} \left\{ (1-\rho) SA^2S^{\top} + \sigma_w^2 I\right\}^{-1}.
\]

\subsection{LASSO estimation}
If $\rho$ is high, $b$ becomes sparse,
and thus we can consider the following optimization problem:
\[
 b_{\mathrm{LASSO}} = \argmin_{x \in \mathbb{R}^N} \lambda \|y - SAx\|^2_2 + \|x\|_1
\]
with a variable $x$,
where $\|\cdot \|_1$ is the $\ell^1$ norm of the vector,
 $\|\cdot \|_2$ is the $\ell^2$ norm of the vector,
and $\lambda$ is a positive number.
LASSO employs the optimal solution of the problem as the estimated signal for the transmitted signal $b$.

\section{Proposed MAP-SOAV Optimization Scheme}
\label{sec:alg}
\subsection{MAP-SOAV estimation}
MAP estimation is to choose the vector $b_{\mathrm{MAP}}$ that maximizes \textit{a posteriori} probability based on the given measurement $y$:
\[
 b_{\mathrm{MAP}} = \argmax_{x \in \mathcal{A}^N} P(x|y)
\]
with a variable $x = [x_1,\ldots,x_N]^{\top}$, where $P(x|y)$ is the posterior probability.
From the Bayes' theorem,
\[
 \begin{array}{ccl}
 \displaystyle{\argmax_{x \in \mathcal{A}^N} P(x|y)} & = & \displaystyle{\argmax_{x \in \mathcal{A}^N} \dfrac{P(y|x) P(x)}{P(y)}} \\
  & = & \displaystyle{\argmax_{x \in \mathcal{A}^N} P(y|x)P(x)} \\
  & = & \displaystyle{\argmin_{x \in \mathcal{A}^N} \{ -\log P(y|x)-\log P(x)\}}. \\
 \end{array}
\]
Here,
\[
 \begin{array}{ccl}
 -\log P(y|x) & = & - \log \left\{ \dfrac{1}{\sqrt{2\pi \sigma_w^2}} \mathrm{exp} \left(-\dfrac{\|y-SAx\|_2^2}{2\sigma_w^2}\right) \right\} \\
  & = & \dfrac{1}{2\sigma_w^2} \|y- SAx\|_2^2 + \dfrac{1}{2} \log (2\pi \sigma_w^2). \\
 \end{array}
\]
Let $r_0,\ldots,r_L\ (r_0<\ldots<r_L)$ be the candidates of transmission symbols
and $p_l$ be the probability with $r_l$.
For any fixed $n\ (1\leq n \leq N)$ and
$r_i\ (i=0,\ldots,L)$,
\[
 \begin{array}{ccl}
 P(x_n = r_i) & = & p_0^{\frac{\Pi_{l \neq 0} (r_i-r_l)}{\Pi_{l \neq 0}(r_0 - r_l) }} \cdots p_L^{\frac{\Pi_{l \neq L} (r_i-r_l)}{\Pi_{l \neq L}(r_L - r_l)}} \\
  & = & \displaystyle{\Pi_{j=0}^L p_j^{\overline{r}_{ij}}}, \\
 \end{array}
\]
where
\[
 \overline{r}_{ij} \triangleq \dfrac{\Pi_{l \neq j} (r_i - r_l) }{\Pi_{l \neq j} (r_j - r_l)}.
\]
Therefore, for any $x \in \{r_0,\ldots,r_L\}^N$,
\[
 \begin{array}{ccl} \vspace{.3em}
 P(x) & = & \displaystyle{ \Pi_{n=1}^N \Pi_{j=0}^L} p_j^{ \overline{x}_{nj} }  \\ \vspace{.3em}
  & = & \displaystyle{ \Pi_{j=0}^L p_j^{ (\overline{x}_{1j} + \ldots + \overline{x}_{Nj} ) }   } \\
  & = & \displaystyle{ \Pi_{j=0}^L p_j^{N - \|x - r_i {\bf 1}_N\|_0 }  }, \\
 \end{array}
\]
where $\| \cdot \|_0$ denotes the $\ell^0$ pseudo-norm,
\[
 \overline{x}_{nj} \triangleq \dfrac{\Pi_{l \neq j} (x_n - r_l)}{\Pi_{l \neq j} (r_j - r_l)},
\]
and ${\bf 1}_N$ is defined by the $N$ dimension vector whose elements are all $1$.
Hence,
\[
 \begin{array}{ccl}
 - \log P(x) & = & \displaystyle{- \sum_{l=0}^L (N - \|x -r_l {\bf 1}_N \|_0 ) \log p_l} \\
  & = & \displaystyle{\sum_{l=0}^L (\log p_l) \|x-r_l {\bf 1}_N \|_0  - N\sum_{l=0}^L \log p_l}.\\
 \end{array}
\]
Summarizing the above calculations, we obtain the following relationship:
\begin{equation}
 \begin{array}{ccl}
 b_{\mathrm{MAP}} & = & \displaystyle{ \argmin_{x \in \mathcal{A}^N} \Big\{ \frac{1}{2\sigma_w^2} \|y - SAx\|_2^2 + \frac{1}{2} \log (2\pi \sigma_w^2)} \\
 & &  \displaystyle{+ \sum_{l=0}^L (\log p_l) \|x - r_l {\bf 1}_N \|_0 - N\sum_{l=0}^L \log p_l \Big\}} \\
 & = & \displaystyle{ \argmin_{x \in \mathcal{A}^N} \Big\{ \frac{1}{2\sigma_w^2} \|y - SAx\|_2^2} \\
 & & \displaystyle{\hspace{5em} + \sum_{l=0}^L (\log p_l) \|x - r_l {\bf 1}_N \|_0 \Big\}}. \\
 \end{array}
\label{eq:summ}
\end{equation}
We get the MAP vector by solving the optimization problem.
However, it is difficult to solve (\ref{eq:summ}) because it becomes a combinatorial optimization problem due to the finiteness of $\mathcal{A}$ and the $\ell^0$ pseudo-norms.

\begin{figure}[t]
\centering
\includegraphics[width = .98\linewidth]{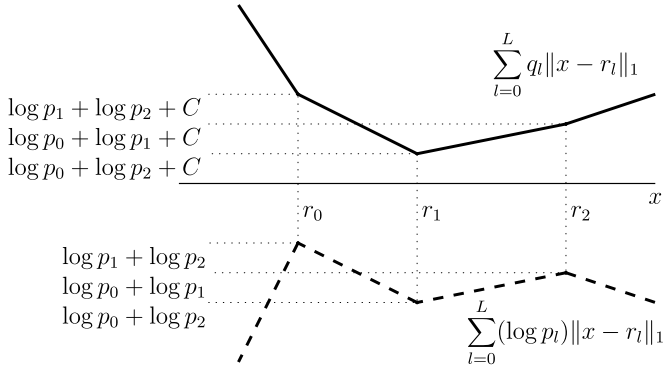}
\caption{Convex relaxation example for $N=1$ and $L=2$.}
\label{relax}
\end{figure}

To tackle with the difficulty, we consider a convex relaxation problem of (\ref{eq:summ}).
First, we expand the domain $\mathcal{A}^N$ into $\mathbb{R}^N$.
That is, we consider
\[
 \begin{array}{ccl}
 b'_{\mathrm{MAP}} & \triangleq & \displaystyle{ \argmin_{x \in \mathbb{R}^N} \Big\{ \frac{1}{2\sigma_w^2} \|y - SAx\|_2^2} \\
 & & \displaystyle{\hspace{5em} + \sum_{l=0}^L (\log p_l) \|x - r_l {\bf 1}_N \|_0 \Big\}}. \\
 \end{array}
\]
Next, replacing the $\ell^0$ pseudo-norms with $\ell^1$ norms based on the idea of the SOAV optimization~\cite{DSR}, we define
\[
 \begin{array}{ccl}
 b''_{\mathrm{MAP}} & \triangleq & \displaystyle{ \argmin_{x \in \mathbb{R}^N} \Big\{ \frac{1}{2\sigma_w^2} \|y - SAx\|_2^2} \\
 & & \displaystyle{\hspace{5em} + \sum_{l=0}^L (\log p_l) \|x - r_l {\bf 1}_N \|_1 \Big\}}. \\
 \end{array}
\]
The problem is not a combinatorial problem,
while it is also not a convex optimization problem due to the fact that $p_l$ is less than 1.
Thus, we further consider to relax
\begin{equation}
 \displaystyle{ \frac{1}{2\sigma_w^2}\|y -SAx\|_2^2 + \sum_{l=0}^L (\log p_l) \|x - r_l {\bf 1}_N\|_0 + C,}
 \label{eq11}
\end{equation}
with
\begin{equation}
 \displaystyle{ \frac{1}{2\sigma_w^2}\|y-SAx\|_2^2 + \sum_{l=0}^L q_l \|x - r_l {\bf 1}_N \|_1}
 \label{eq12}
\end{equation}
for a sufficiently large constant $C \in \mathbb{R}$.
Note that (\ref{eq12}) can be regarded as a relaxation of (\ref{eq11}) when the value of (\ref{eq12}) is equal to the value of (\ref{eq11}) for any $x \in \{r_0,\ldots,r_L\}^N$.
See Fig.~\ref{relax} that illustrates an example of the relaxation for $N=1$ and $L=2$.

We consider to derive the condition of $q_l$ for the relaxation.
Since we have
\[
 \displaystyle{ \sum_{l=0}^L (\log p_l) \|r_i-r_l\|_0 + C = \sum_{l \neq i} (\log p_l) + C,}
\]
if $x_n$ is equal to $r_i$, the condition of the relaxation is given by
\[
 \displaystyle{ \sum_{l \neq i} q_l |r_i -r_l| = \sum_{l \neq i} (\log p_l) + C}
\]
for any $i$.
This condition is equivalent to the following linear equation:
\[
 Rq = P_C,
\]
where
\[
 \begin{array}{l}
 R \triangleq \left[
 \begin{array}{ccccc}
 0 & |r_1 - r_0| & |r_2-r_0| & \cdots & |r_L - r_0| \\
 |r_0-r_1| & 0 & |r_2-r_1|  & \cdots & |r_L - r_1| \\
 \vdots &  & \ddots & & \\
 |r_0 - r_L| &  & & & 0 \\
 \end{array}\right], \\
 q \triangleq [q_0, \ldots, q_L]^{\top}, \\
 P_C \triangleq \displaystyle{\left[  \sum_{l \neq 0} ( \log p_l) + C, \ldots, \sum_{l \neq L} (\log p_l) + C \right] ^{\top}}. \\
 \end{array}
\]
We can find an appropriate $q$ by solving the equation.

With the obtained $q$, we consider the following optimization problem:
\begin{equation}
 \begin{array}{l}
 b_{\mathrm{MAP-SOAV}} \\
 \triangleq \displaystyle{ \argmin_{x \in \mathbb{R}^N} \left( \frac{1}{2\sigma_w^2}\|y-SAx\|_2^2 + \sum_{l=0}^L q_l \|x - r_l {\bf 1}_N \|_1\right) }. \\
 \end{array}
 \label{opt}
\end{equation}
We call the estimation method the MAP-SOAV method.

\subsection{Algorithm based on a proximal splitting method}
Next, we consider to apply a proximal splitting algorithm that is known as a fast algorithm solving convex optimization problems.
We give the closed form of the proximity operator for the algorithm when we employ the binary phase shift keying (BPSK) modulation.
Note that, $L=2$, $r_0 = -1$, $r_1 = 0$, $r_2 = 1$, and $p_1=\rho,p_0= p_2 = (1-\rho)/2$ for standard BPSK modulation.

Letting
\[
 f(x) \triangleq \frac{1}{2\sigma_w^2} \|y - SAx\|_2^2
\]
and
\[
 g(x) \triangleq \sum_{l=0}^L q_l \|x-r_l {\bf 1}_N\|_1,
\]
we rewrite the problem as
\[
 \min_{x \in \mathbb{R}^N} \left\{f(x) + g(x)\right\}.
\]
Define the proximity operator of $g$ by
\[
 \prox{g}{x} \triangleq \argmin_{u \in \mathbb{R}^N} \{ g(u) +\frac{1}{2\gamma}\|x-u\|_2^2\}.
\]
Then we have the following proposition.

\begin{figure}[t]
\centering
\includegraphics[width = .98\linewidth]{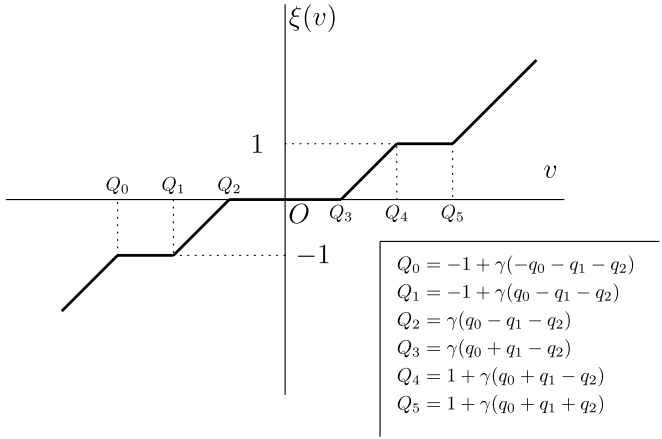}
\caption{Function $\xi(v)$ for $\prox{g}{x}$.}
\label{prox}
\end{figure}

\begin{proposition}
 Let $\xi:\mathbb{R} \to \mathbb{R}$ be
\[
 \xi(v) \triangleq \left\{
 \begin{array}{ll}
 v-\gamma(-q_0-q_1-q_2) & \mathrm{if}\ v < Q_0, \\
 -1& \mathrm{if}\ Q_0 \leq v < Q_1, \\
 v - \gamma(q_0-q_1-q_2) & \mathrm{if}\ Q_1 \leq v < Q_2, \\
 0 & \mathrm{if}\ Q_2 \leq v < Q_3, \\
 v - \gamma(q_0+q_1-q_2) & \mathrm{if}\ Q_3 \leq v < Q_4, \\
 1 & \mathrm{if}\ Q_4 \leq v < Q_5, \\
 v-\gamma(q_0+q_1+q_2) & \mathrm{if}\ Q_5 \leq v, \\
 \end{array}\right.
\]
where
\[
 \begin{array}{l}
 Q_0 \triangleq -1+\gamma(-q_0-q_1-q_2), \\
 Q_1 \triangleq -1 + \gamma(q_0-q_1-q_2), \\
 Q_2 \triangleq \gamma(q_0-q_1-q_2), \\
 Q_3 \triangleq \gamma(q_0+q_1-q_2), \\
 Q_4 \triangleq 1 + \gamma(q_0+q_1-q_2), \\
 Q_5 \triangleq 1 + \gamma(q_0+q_1+q_2), \\
 Q_6 \triangleq 1+\gamma(q_0+q_1+q_2), \\
 \end{array}
\]
(see also Fig.~\ref{prox}).
Then, we have
\[
 \prox{g}{x} = [\xi(x_1), \ldots,\xi(x_N)]^{\top},
\]
where $x_i$ is the $i$th element of $x$.
\end{proposition}
Because this proposition can be derived by straightforward calculations,
the proof is omitted.

With the proximity operator, the following algorithm is the proximal splitting method to solve (\ref{opt}):
\begin{algorithm}
Fix $\tilde{x}^{(1)} \in \mathbb{R}^N,t_1=1$, and $L \in \mathbb{R}$ which is greater than or equal to a Lipshitz constant of $\nabla f(x) = \frac{1}{\sigma_w^2} (SA)^{\top}(SAx-y)$.
For $k \geq 1$,
\[
 \left\{
 \begin{array}{rcl} \vspace{.2em}
 x^{(k)} & = & \displaystyle{\prox{\frac{1}{L}g}{\tilde{x}^{(k)}-\frac{1}{L\sigma_w^2} (SA)^{\top}(SA\tilde{x}^{(k)}-y)}}, \\ \vspace{.2em}
 t_{k+1} & = & \displaystyle{\frac{1+\sqrt{1+4t_{k}^2}}{2},} \\
 \tilde{x}^{(k)} & = & \displaystyle{x^{(k)} + \left( \frac{t_k-1}{t_{k+1}} \right) \left(x^{(k)} - x^{(k-1)}\right).} \\
 \end{array}
 \right.
\]
\end{algorithm}
Here, the following proposition holds~\cite{Fixed-Point}.
\begin{proposition}
If the optimization problem is convex, then the algorithm converges to a solution of the optimization problem.
Moreover, the convergence rate is $\mathcal{O}(1/k^2)$.
\end{proposition}

Finally, we define the following threshold function $\phi_{\alpha}: \mathbb{R} \to \mathbb{R}$,
\[
 \phi (v) \triangleq \left\{
 \begin{array}{ll}
 -1 & \mathrm{if}\ v < -\alpha, \\
 0 & \mathrm{if}\ -\alpha \leq v < \alpha, \\
 1 & \mathrm{if}\ \alpha \leq v, \\
 \end{array}\right.
\]
with a positive number $\alpha$.
With the threshold function, we define $\Phi: \mathbb{R}^N \to \mathbb{R}^N$ by
\[
 \Phi \triangleq [\phi(\cdot),\ldots,\phi(\cdot)]^{\top}
\]
and use
\[
 \hat{b} \triangleq \Phi \left( b_{\mathrm{MAP-SOAV}}\right)
\]
as the estimation signal of the transmitted signal $b$.

\section{Numerical Examples}
\label{sec:sim}
We present some simulation results to illustrate the effectiveness of the proposed method.

In this study, signal-to-noise ratio is defined by
\[
 \begin{array}{ccl}
 \mathrm{SNR} & \triangleq & 10 \log_{10} \left\{ \dfrac{\mathrm{tr}[ \Sigma_b ]}{\mathrm{tr}[ \Sigma_w]}\right\} \\
 & = & 10 \log_{10} \left\{ \dfrac{N(1-\rho)}{M\sigma_w^2} \right\}. \\
 \end{array}
\]
Hence, $\sigma_w^2$ is determined as follows:
\[
 \begin{array}{l}
 \sigma_w^2 = \dfrac{N(1- \rho)}{M} 10^{-\mathrm{SNR}/10}. \\
 \end{array}
\]

In all simulations, the number of users $N$ is set to be $100$
and the number of measurement $M$ is $70$.
Then, $S$ is an element in $\mathbb{R}^{70 \times 100}$.
We employ BPSK modulation,
that is,
\[
 \mathcal{A} = \{1,0,-1\}
\]
and $b \in \mathcal{A}^N$ is generated with the probability 
\[
 \begin{array}{l}
 P(1) = P(-1) = \dfrac{1-\rho}{2}, \\
 P(0) = \rho. \\
 \end{array}
\]
Each component of $S$ is determined by normal Gaussian distribution (choice of $S$ is discussed in~\cite{Xie2013}).
The channel gains
\[
 A = I.
\]
The weight of the $\ell^2$ term $\lambda$ of the LASSO method is $3 \times 10^{1}$.
The threshold value $\alpha$ for the threshold function $\phi_{\alpha}$ is set to be $0.5$.
On each simulation, we evaluate the error ratio by averaging the results obtained in 1000 trials.

\begin{figure}[t]
\centering
\includegraphics[width = .98\linewidth]{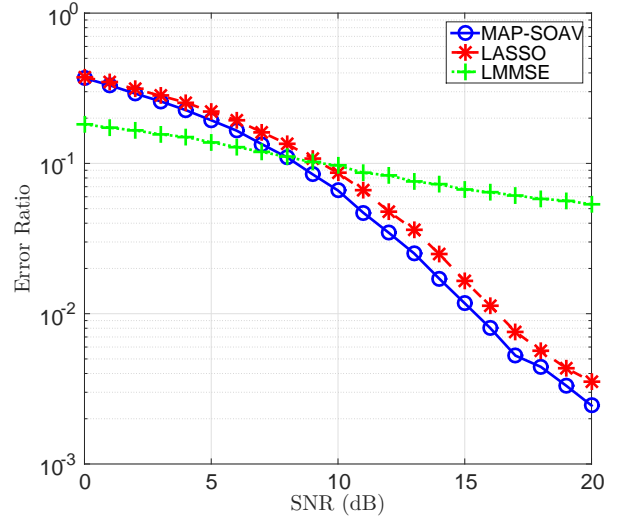}
\caption{SNR vs Error Ratio when the non-active rate $\rho$ is $0.8$.
The user number $N$ is $100$,
the measurement number $M$ is $70$,
and $S$ is generated by normal Gaussian distribution.
(a) The solid line corresponds to MAP-SOAV estimation.
(b) The broken line corresponds to LASSO estimation.
(c) The dotted line corresponds to LMMSE estimation.}
\label{graph01}
\end{figure}

In the first numerical example, the non-active rate $\rho$ is $0.8$.
This setting corresponds to the case that the transmitted signal is sparse.
The constant $C$ is $14.6052$.
It is determined by
\[
 C = \left|\min  \left\{ \sum_{l \neq 0} ( \log p_l), \ldots, \sum_{l \neq L} (\log p_l) \right\}\right| + 5.
\]
Then,
\[
 q = [5,\ 2.0794,\ 5]^{\top}
\]
and the relaxed problem becomes a convex optimization problem.
Fig.~\ref{graph01} shows the SNR vs Error ratio.
Error ratio is defined by the ratio of the number of error components to $N$.
From the figure, we can see that the proposed method is slightly better than the LASSO method and is effective when SNR is high.

\begin{figure}[t]
\centering
\includegraphics[width = .98\linewidth]{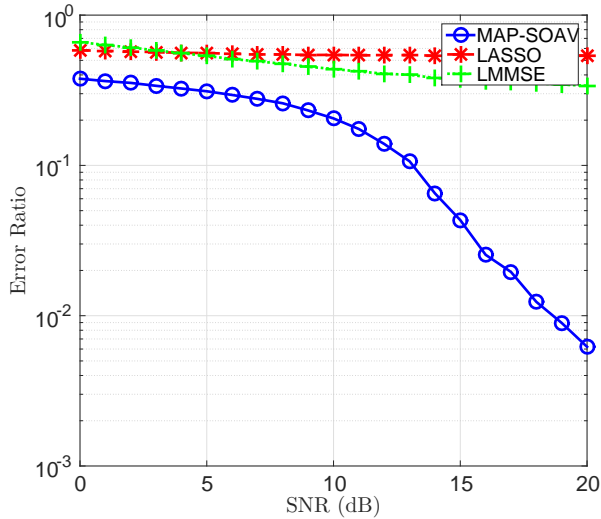}
\caption{SNR vs Error Ratio when the non-active rate $\rho$ is $0.05$.
The user number $N$ is $100$,
the measurement number $M$ is $70$,
and $S$ is generated by normal Gaussian distribution.
(a) The solid line corresponds to MAP-SOAV estimation.
(b) The broken line corresponds to LASSO estimation.
(c) The dotted line corresponds to LMMSE estimation.}
\label{graph02}
\end{figure}

Fig.~\ref{graph02} shows the SNR vs Error Ratio when the non-active rate $\rho$ is $0.05$.
Consequently, the values of most of the symbols get $1$ or $-1$.
The constant $C$ is $13.7402$.
Then,
\[
 q = [6.1256,\ -2.2513,\ 6.1256]^{\top}.
\]
It is notable that optimality of the result of the algorithm is not guaranteed since the optimization problem does not hold convexity in this case.
However, the figure indicates that the performance of MAP-SOAV estimation is much better than LMMSE estimation and LASSO estimation when the non-active rate is low.

\begin{figure}[t]
\centering
\includegraphics[width = .98\linewidth]{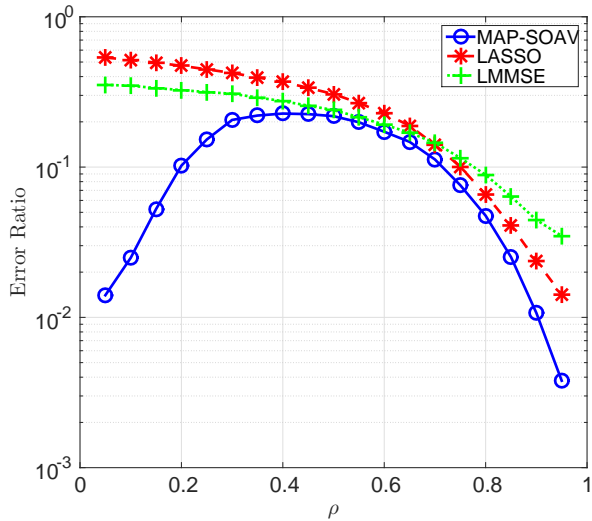}
\caption{The non-active rate $\rho$ vs Error Ratio.
The user number $N$ is $100$,
the measurement number $M$ is $70$,
$\sigma^2_w$ is $0.0226$,
and $S$ is generated by normal Gaussian distribution.
(a) The solid line corresponds to MAP-SOAV estimation.
(b) The broken line corresponds to LASSO estimation.
(c) The dotted line corresponds to LMMSE estimation.}
\label{graph03}
\end{figure}

Finally, we investigate the effect of  $\rho$.
Fig.~\ref{graph03} shows the non-active rate vs Error Ratio.
In this simulation, the variance of the noise is determined by
\[
 \begin{array}{ccl}
 \sigma_w^2 & = & 0.05 \times 10^{-5/10} \times N/M \\
  & = & 0.0226. \\
 \end{array}
\]
Note that the SNR defined above changes with the variation of $\rho$.
From the figure, it can be seen that the MAP-SOAV method has the best performance .
In particular, when the closer $\rho$ is to $0$ or $1$, the more the proposed scheme is effective.

\section{Conclusion}
\label{sec:conc}

In this article,
we have proposed a new multiuser detection method taking account of discreteness as prior information of transmitted signals in the uplink M2M communications with a centralized structure.
The synchronous multiuser communication system model with CDMA has been given for an AWGN channel.
The transmitted symbols have been related to the received signal by a under-determined linear equation with white Gaussian noise.
We have formulated the detection problem as the MAP estimation, which is relaxed to a convex optimization called the SOAV optimization.
We have given the proximity operator in a closed form for a proximal splitting algorithm that solves the SOAV optimization problem efficiently.
Numerical simulations have been shown to illustrate the effectiveness of the proposed approach compared with the LMMSE and LASSO method.
It has been shown that the proposed approach is better than the other method in most cases.
In particular, when the non-active rate is close to $0$ or $1$, the proposed scheme is very effective.






%


\bibliographystyle{IEEEtran}
\bibliography{sshrrefs}

\end{document}